\documentclass[aps,twocolumn,showpacs,preprintnumbers,amsmath,amssymb,nofootinbib,superscriptaddress,showkeys]{revtex4}

\usepackage{epsfig}
\usepackage{graphicx}

\def\slashchar#1{\setbox0=\hbox{$#1$}
   \dimen0=\wd0 \setbox1=\hbox{/} \dimen1=\wd1
   \ifdim\dimen0>\dimen1 \rlap{\hbox to \dimen0{\hfil/\hfil}} #1
   \else  \rlap{\hbox to \dimen1{\hfil$#1$\hfil}} / \fi}

\begin{document}

\title{{Pion electromagnetic form factor, perturbative QCD, and 
large-$N_c$ Regge models}\footnote{Supported by Polish Ministry of Science
and Higher Education, grant N202~034~32/0918, Spanish
DGI and FEDER funds with grant FIS2005-00810, Junta de Andaluc{\'\i}a
grant FQM225-05, and EU Integrated Infrastructure Initiative Hadron
Physics Project contract RII3-CT-2004-506078.  }}
\author{Enrique Ruiz Arriola}\email{earriola@ugr.es} 
\affiliation{Departamento de F\'{\i}sica At\'omica, Molecular y Nuclear, Universidad de Granada, E-18071 Granada, Spain}
\author{Wojciech Broniowski} \email{Wojciech.Broniowski@ifj.edu.pl}
\affiliation{The H. Niewodnicza\'nski Institute of Nuclear Physics, PL-31342~Krak\'ow, Poland} 
\affiliation{Institute of Physics, Jan Kochanowski University, PL-25406~Kielce, Poland}

\date{21 July 2008}

\begin{abstract} 
We present a construction of the pion electromagnetic form factor
where the transition from large-$N_c$ Regge vector meson dominance
models with infinitely many resonances to perturbative QCD is built in explicitly.
The construction is based on an appropriate assignment of residues to the Regge poles, which 
fulfills the constraints of the parton-hadron duality and perturbative QCD.
The model contains a slowly falling off non-perturbative contribution which
dominates over the perturbative QCD radiative corrections for the
experimentally accessible  momenta. The leading order and next-to-leading order 
calculations show a converging pattern
which describes the available data within uncertainties, while the
onset of asymptotic QCD takes place at extremely high momenta, 
$Q \sim 10^3-10^4 {\rm GeV}$. The method can be straightforwardly extended to study 
other form factors where the perturbative QCD result is available.  
\end{abstract}

\pacs{12.38.Lg, 11.30, 12.38.-t} 

\keywords{Pion electromagnetic form factor, large $N_c$ Regge models, perturbative QCD}

\maketitle

\section{Introduction}

The composite nature of hadrons can be best seen by studying their
electromagnetic form factors at a sufficiently large momentum
transfer~\cite{Gourdin:1974iq}. The pion, being the lowest $u$ and $d$
quark-antiquark excitation of the vacuum and identified with the
would-be Goldstone massless mode of the spontaneously broken chiral symmetry,
provides a simplest candidate to test our present knowledge
on hadronic interactions. Due to relativistic and gauge invariance the
pion charge form factor (we take $\pi^+$ for definiteness) can be written as
\begin{eqnarray}
\langle \pi^+ (p') | J_\mu^{\rm em} (0) | \pi^+ (p) \rangle = 
\left(p'^\mu + p^\mu \right) F(q^2)     
\end{eqnarray} 
with $q=p'-p$ and $ J_\mu^{\rm em} (x) = \sum_{q=u,d,s, \dots} e_q\bar
q (x) \gamma_\mu q(x)$ is the electromagnetic current, with $e_q$ 
denoting the quark charge in units of the elementary charge. 
The charge normalization requires
\begin{eqnarray}
F(0)= 1.
\end{eqnarray} 
The pion charge form factor has been the subject of intense experimental
efforts~\cite{Bebek:1974ww,Bebek:1977pe,Amendolia:1983di,Amendolia:1986wj,Volmer:2000ek,Horn:2006tm,Tadevosyan:2007yd,Blok:2002ew}.
Moreover, it is expected to be measured at TJLAB in the space-like
range of $1~{\rm GeV}^2 \le - t\le 6~{\rm GeV}^2$ with unprecedented high
precision $\Delta (-tF(t)) \sim 0.02~{\rm GeV}^2$. The results might be
used as a stringent test of the perturbative QCD (pQCD) radiative corrections. Actually, in the 
space-like region where $t=-Q^2$, $F(t)$ is real and at large $Q^2$
values the pQCD methods can be applied, yielding
asymptotically~\cite{Brodsky:1973kr,Brodsky:1974vy,Farrar:1979aw,Radyushkin:1977gp,Efremov:1978rn,Efremov:1979qk}
\begin{eqnarray}
F(-Q^2) &=& \frac{16 \pi f_\pi^2 \alpha(Q^2)}{Q^2}\left[1+ 6.58
  \frac{\alpha (Q^2)}{\pi} + \dots \right], \nonumber 
\\ \qquad && Q^2 \gg M^2 \label{pQCDff}
\end{eqnarray} 
with $f_\pi=92.3 {\rm MeV}$ denoting the pion weak decay constant, and
$M$ the lowest vector meson mass.  Further higher-order power corrections are
of the order ${\cal O} (1/Q^4)$ and correspond to higher twist
operators~\cite {Braun:1999uj,Agaev:2005gu}. The form factor depends
logarithmically on the scale through the running coupling
constant
\begin{eqnarray}
&&\alpha(Q^2)=\frac{4\pi}{\beta_0 \log(Q^2/\Lambda^2)} \\
&&\beta_0=\frac{11}{3}N_c-\frac{2}{3}N_f. \label{gambe}
\end{eqnarray}
We use the $\overline{\rm MS}$ scheme and the
factorization scale coinciding with the renormalization scale. Also,
the asymptotic form of the pion parton distribution amplitude,
$\phi_\pi(x)=6x(1-x)$, is used. Details of the complete analysis
may be found in Ref.~\cite{Melic:1998qr}.
The second term in brackets in Eq.~(\ref{pQCDff}) is the
next-to-leading (NLO) correction. It is at an acceptable $20\%$ level
when $\alpha \sim 0.1$, which suggests that one might observe this
radiative correction at relatively large scales, $Q^2 \sim
M_Z^2$. 

In the intermediate energy region the form factor
behaves to a very good accuracy as
\begin{eqnarray}
F(-Q^2) = \frac{M_V^2}{Q^2+ M_V^2}   \, , \qquad Q^2 \leq M_V^2 
\label{eq:vmd}
\end{eqnarray} 
with $M_V = 720 {\rm MeV}$, complying to the old vector meson
dominance models (VMD) (see e.g. Ref.~\cite{Sakurai:1969} and
references therein) and showing no obvious trace of the pQCD behavior.
In the region close to the zero momentum transfer chiral corrections
become important~\cite{Gasser:1984ux,Leutwyler:2002hm}. For time-like
momenta the pion form factor becomes complex and can be related by
crossing to the $e^+ e^- \to \pi^+ \pi^-$ annihilation amplitude,
$\langle \pi^+ \pi^- | J_\mu^{\rm em} | 0 \rangle = F(s) (p^\mu +
p'^\mu)$, where the final state interactions due to $\pi\pi$
scattering and unitarity play a crucial
role~\cite{Nieves:1999bx}. While both the time-like and the space-like
regions are related by an unsubtracted dispersion
relation~\cite{Donoghue:1996bt},
\begin{eqnarray}
F(t) = \frac1\pi \int_{t_0}^\infty \frac{{\rm Im} F(t')}{t'-t-i\epsilon} dt'  ,
\label{eq:dr}
\end{eqnarray} 
the well-known time-like region does not determine unambiguously when
the onset of the pQCD takes place.  Actually, the single VMD model
shows that even in the space-like region as low as $Q \sim m_\rho$ the
traces of chiral-logs and final state interactions are meager.

Given the fact that the pQCD effects cannot directly be observed at
presently available energies, numerous phenomenological QCD-based
approaches and model calculations have been suggested in order to
understand the transition from the soft to hard scales. They include
standard QCD sum rules~\cite{Ioffe:1982ia}, local-duality QCD sum
rules~\cite{Nesterenko:1982gc,Braguta:2007fj}, light-cone QCD sum
rules~\cite{Braun:1994ij}, nonlocal
condensates~\cite{Bakulev:1991ps,Bakulev:2004cu}, Schwinger-Dyson
equations~\cite{Maris:1998hc}, instanton-based
models~\cite{Faccioli:2002jd, Nam:2007gf}, constituent quark
models~\cite{Cardarelli:1995hn}, nonlocal quark
models~\cite{Pagels:1979hd,Ito:1991pv}, {\em etc.}  The scale of the
onset of pQCD provoked heated debates in the past. The problem is
crucial, as it provides a decisive finger print of the underlying
quark-gluon substructure of the pion.  We note that the upcoming
lattice QCD calculations extending the work reported
in~\cite{Bonnet:2004fr,Hashimoto:2005am,Brommel:2005ee,Brommel:2006ww,Hsu:2007ai}
can directly verify this issue without necessarily spanning such a
wide energy window as in the experiment. The reason is that a lot of
progress has been achieved in extrapolating the lattice data to the
chiral limit, which incorporates the enhancement and nonlinearities
triggered by the chiral logs.

The class of calculations listed above contains quarks and gluons as
explicit dynamical degrees of freedom, and hence requires a detailed knowledge of
the pion wave function. On the other hand the parton-hadron duality
implies that any hadronic property be describable in the purely hadronic language without an explicit
reference to the basic fundamental fields. For instance, the success of the simple VMD fit
for the pion charge form factor suggests the inclusion of further radially
excited $I^G J^{PC}= 1^+ 1^{--}$ states, $\rho', \rho'', \rho'''
\dots$, 
\begin{eqnarray}
F(t) = \sum_{V=\rho,\rho',\dots}^{V_{\rm max}} \frac{c_V M_V^2}{M_V^2-t} . \label{sumrho} 
\end{eqnarray} 
This finite sum involves states with a mass below $M_{V_{\rm max}}$,
the highest allowed vector meson mass which acts as a high energy
cut-off. Thus, it could reliably reproduce the data (see below) in a
region where $Q^2 < M_{V_{\rm max}}^2$, and will only produce inverse
integer powers of $Q^2$ asymptotically when $Q^2 \gg M_{V, {\rm
max}}^2$. This is in formal contradiction with Eq.~(\ref{pQCDff}),
where there is no high energy cut-off and the behavior $1/(Q^2 \log
Q^2)$ is obtained. Thus, infinitely many states are clearly needed.
This complies to the t'Hooft large-$N_c$ limit~\cite{tHooft:1973jz},
where any hadronic amplitude can be written in terms of tree diagrams
with (infinitely many) mesons and glueballs.  In particular, in the
large-$N_c$ limit the pion form factor can be written in the form
(\ref{sumrho}) with {\em infinitely many} resonances.

Based on the success of the Veneziano-Lovelace-Shapiro dual resonance
model (see e.g. \cite{Veneziano:1974dr,Mandelstam:1974fq} and
references therein) Suura~\cite{PhysRevLett.23.551} and
Frampton~\cite{Frampton:1969ry} proposed analytic models which have
recently been resurrected and further elaborated by
Dominguez~\cite{Dominguez:2001zu,Dominguez:2008gg}.  Incidentally, the
resulting expressions for the pion charge form factor turn out to be
quite similar to the AdS/CFT hard-wall and soft-wall calculations
carried out
in~\cite{Brodsky:2007hb,Grigoryan:2007wn,Kwee:2007nq}. Despite the
successful fit to the data, these calculations do not reproduce the
formal asymptotic pQCD behavior, a fact which has been interpreted as
an intrinsic limitation of the approach~\cite{Grigoryan:2007wn}. This
poses an intriguing puzzle: how do hadronic large-$N_c$ models satisfy
the QCD constraints, including the presence of logarithms?  Quite
generally, pQCD predicts integer powers and logarithms of $Q^2$,
whereas the models of
Refs.~\cite{PhysRevLett.23.551,Frampton:1969ry,Dominguez:2001zu,Dominguez:2008gg}
are able to generate fractional powers.

In the present paper we analyze the problem for the case of the pion
charge form factor and show how the pQCD constraints can judiciously
be implemented in a large-$N_c$ Regge model in an exact manner and at
the same time preserve the good description of the experimental
data. The essence of the approach is a careful assignment of coupling
constants to the infinitely many resonances. As a result, the form
(\ref{pQCDff}) emerges from the infinite sum (\ref{sumrho}).  We term
the mechanism the {\em power-to-log transmutation}, which essentially
corresponds to a suitable superposition of fractional twist operators
in the Regge model of
Refs.~\cite{PhysRevLett.23.551,Frampton:1969ry,Dominguez:2001zu,Dominguez:2008gg}
.
The present study follows our investigation of the two-point 
functions~\cite{RuizArriola:2006gq,Arriola:2006sv}. In a previous
paper~\cite{RuizArriola:2006ii} we have shown how the large-$N_c$ Regge
models can be used to deal with the $\gamma^* \pi^0 \to \gamma$
transition form factor, where the radiative pQCD corrections
characterized by the relevant anomalous dimensions are generated with the suitable
QCD evolution equations.

\section{Meson dominance}

In this preparatory Section we introduce the basic definitions and
notation for the pion form factor in VMD models.
The electromagnetic current is written as $J^{\mu, {\rm em }} (x) = B^\mu (x)/2+
J_V^{\mu,3} (x)$ with $B^\mu (x) = \sum_q \bar q(x) \gamma^\mu q(x) /N_c$
being the baryon current and $J_V^{\mu a} (x) = \sum_q \bar q(x) \tau^a
\gamma^\mu q(x) /2$ the iso-vector current. Using
the isospin invariance, assumed throughout, we have
\begin{eqnarray}
\langle \pi^a (p')| J_V^{\mu, b} (0) | \pi^c (p)\rangle = \epsilon^{abc}
\left(p'^\mu + p^\mu \right) F(q^2) \, , 
\end{eqnarray}
with $| \pi^i (p) \rangle $ denoting a pion state, and $a$, $b$, $c$ the Cartesian isospin
indices. In the large-$N_c$ limit the {\em meson dominance} of the pion
charge form factor is the statement that one can parameterize the 
(iso-vector) current as a superposition of vector meson fields, $\rho_{n,\mu}^a(x)$,
\begin{eqnarray}
J_V^{\mu, a} (x) = \sum_n F_V (n) M_V (n) \rho_n^{\mu, a} (x), \label{jv}
\end{eqnarray} 
where $n=0$  corresponds to the  ground state $\rho (770)$  meson, and
higher values  of $n$ to excited states.   Correspondingly, the matrix
element between the vacuum and the one-vector meson state is
\begin{eqnarray}
\langle 0 | J_V^{a \mu} (0) | \rho_n^b , \epsilon \rangle = \delta^{ab}
M_V (n) F_V (n) \epsilon^\mu ,
\end{eqnarray} 
with $\epsilon_\mu$ denoting the vector-meson
polarization. The coupling constants may be determined from the
electromagnetic decay $\rho_n \to e^+ e^- $ using the formula 
\begin{eqnarray}
\Gamma (\rho_n \to e^+ e^- )= \frac{4 \pi \alpha^2}{3}
\frac{F_V (n)^2}{M_V (n)}, 
\end{eqnarray}
for the partial decay rate. For $M_V=m_\rho= 770 {\rm MeV} $ and
$\Gamma (\rho \to e^+ e^- )= 6.5 {\rm keV}$ one gets $F_V =F_\rho =
150 {\rm MeV}$.

The two-point vector current-vector current correlator is defined as
\begin{eqnarray}
\Pi_{V}^{\mu a, \nu b} (q) &=& i \int d^4 x e^{-i q \cdot x} \langle
0 | T \left\{ J_V^{\mu a} (x) J_V^{\nu b} (0) \right\} | 0 \rangle
\nonumber \\ &=& \Pi_V (q^2) \, \left( q^\mu q^\nu - g^{\mu \nu} q^2 \right)
\delta^{ab}, \label{corel} 
\end{eqnarray}  
where 
\begin{eqnarray}
\Pi_V (t) =  \sum_n
\frac{F_V(n)^2}{M_V(n)^2-t}. \label{VV}
\end{eqnarray} 
The quark-hadron duality for large values of $t$ in (\ref{VV}) requires the Regge model -- parton
model matching condition~\cite{RuizArriola:2006gq,Arriola:2006sv} for
asymptotically large values of the radial quantum number $n$,
\begin{eqnarray}
\frac{F_V (n)^2}{d M_V^2 (n) /dn}  \sim \frac{N_c}{24 \pi^2} \, . 
\label{eq:regge-dens}
\end{eqnarray} 
For the radial Regge model (see next Section) $d M_V^2 (n) /dn=a={\rm const.}$, hence at large $n$ we must
have $F_V(n)={\rm const.}$~\cite{RuizArriola:2006gq,Arriola:2006sv}.

The vector meson-pion-pion amplitude is 
\begin{eqnarray}
 \langle \pi^a (p') | \rho_{n,\mu}^b (0) | \pi^c (p) \rangle =
 (p'_\mu+p_\mu) \frac{ \epsilon^{abc} g_{V\pi \pi}(n)}{M_V(n)^2-t} \, ,
\end{eqnarray} 
with $ g_{V\pi \pi}(n)$ the coupling constant. This yields 
the $\rho_n \to \pi \pi$ partial decay rate
\begin{eqnarray}
\Gamma(\rho_n \to \pi \pi) = \frac{g_{V\pi\pi}^2 M_V}{48 \pi} \left(
1- \frac{4 m_\pi^2}{M_V^2} \right)^\frac32 \, .
\end{eqnarray}
For the $\rho (770)$ meson one gets $g_{\rho \pi \pi} \simeq6$ for
$\Gamma (\rho \to 2 \pi)=150 {\rm MeV}$.

For the pion electromagnetic form factor we have 
\begin{eqnarray}
F(t) &=& \sum_n  \frac{c_n M_V(n)^2}{M_V(n)^2-t} \, \nonumber \\
c_n&=& \frac{F_V(n) g_{V\pi\pi} (n)}{M_V(n)}.  \label{ffv}
\end{eqnarray} 
With the adopted conventions we note that $F_V(n)$ has the
dimension of energy, while $g_{V\pi\pi}(n)$ and $c_n$ are dimensionless. 
Note that the signs of the residues appearing in the form factor
(\ref{ffv}) may a priori be positive or negative, while all
contributions to the two-point correlator (\ref{VV}) are positive.
The possibility of different signs in Eq.~(\ref{ffv}) provides a
mechanism for cancellation.  The form factor satisfies the dispersion
relation (\ref{eq:dr}) with the spectral density
 \begin{eqnarray}
\frac1{\pi}{\rm Im} F(t) = \sum_n c_n M_V(n)^2 \, \delta
(t-M_V(n)^2) \, .
\end{eqnarray} 
Note that with the previously listed parameters for the lowest
$\rho(770) $ resonance one has $ c_\rho = g_{\rho \pi \pi} F_\rho
/m_\rho= 1.17 $. Because of charge conservation this requires higher
states with negative $c_n$ coefficients. In fact, taking
Eq.~(\ref{sumrho}) with the physical vector meson masses, $m_\rho =770
{\rm MeV}$, $m_{\rho'}=1459(10) {\rm MeV}$, $m_{\rho''}=1720(20) {\rm
MeV}$ and $m_{\rho'''}=2000(30) {\rm MeV}$ and using the coupling
constants $c_n$ as fit parameters to the electromagnetic form factor
data in the intermediate $Q^2$ range, $ 0.6 {\rm GeV}^2 < Q^2 <
2.4~{\rm GeV}^2$, yields $c_\rho=1.25$, $c_{\rho'}=-0.17$ for two
resonances, $c_\rho=1.39$, $c_{\rho'}=-0.53$, $c_{\rho''}=0.26$ for
three resonances, and $c_\rho=1.39$, and $c_{\rho'}=-0.53$ ,
$c_{\rho''}=0.26$, $c_{\rho'''}=-0.004$ for four resonances.  Such an
approach, although phenomenologically appealing and numerically stable
for the lowest energy states, can only yield an integer power fall-off
and, as already mentioned, does not match to pQCD,
Eq.~(\ref{pQCDff}), at high energies.

Strictly speaking one should consider in Eq.~(\ref{ffv}) the leading
large-$N_c$ contributions to the vector meson parameters. According to
Ref.~\cite{tHooft:1973jz} one has $M_V(n) \sim N_c^0$, $F_V(n) \sim
\sqrt{N_c}$, and $g_{V \pi\pi} (n)\sim 1/\sqrt{N_c}$, such that $F(t)
\sim N_c^0$. Corrections to this behavior are generally $1/N_c$
suppressed relative to the leading order and hence we expect at worse
a $30\%$ detuning of the physical values. The large-$N_c$ dependence
of meson parameters has been studied in unitarized chiral perturbation
theory approaches yielding a larger value for the vector meson mass
when $m_\rho^{N_c}\to m_\rho^\infty \sim 1.2
m_\rho^{N_c=3}$~\cite{Pelaez:2003dy,Pelaez:2006nj}. Chiral quark
models at the one loop level are large $N_c$ motivated. The Spectral
Quark Model~\cite{RuizArriola:2003bs} reproduces by construction the
simple VMD result, Eq.~(\ref{eq:vmd}), for the pion form factor
providing, in addition, the value $m_\rho^2 = 24 \pi^2 f_\pi^2 /N_c$
which for $f_\pi=92.3 {\rm MeV}$ yields $m_\rho \sim 820 {\rm MeV}$, a
larger value than the physical mass. The trend to an increased value
of the $\rho$-meson mass can also be traced when in a fit of the two
resonance version of the generalized VMD, Eq.~(\ref{sumrho}), the
lowest mass state is allowed to vary. Keeping $m_{\rho'}=1460 {\rm
MeV}$ this yields $c_\rho=1.29$, $c_{\rho'}=1-c_\rho=-0.29$ and
$m_\rho=864 {\rm MeV}$.

\section{Regge models}

We now proceed to review the Regge models in the scope necessary for
our analysis, in particular regarding the pion electromagnetic form
factor.
The radial Regge trajectories are 
\begin{eqnarray}
M_n^2 = M^2 + a n. \label{reggep}
\end{eqnarray} 
The slope of the radial Regge trajectory, $a$, may be identified with
the string tension, $\sigma = a/(2\pi)$, which for heavy quarks
corresponds to the confining potential $V(r)= \sigma r$. Acceptable
values are in the range $\sigma = 420-500 {\rm
MeV}$~\cite{Anisovich:2000kxa}.  In this work we use for definiteness
\begin{eqnarray} 
\sigma=450~{\rm MeV},\;\; M = 820~{\rm MeV}. \label{param} 
\end{eqnarray}
As mentioned above, these parameters need not exactly reproduce the
physical values, as the accuracy of the present large-$N_c$ Regge
approach is not expected to be better than the large-$N_c$ expansion
itself. Fortunately, the pion electromagnetic form factor turns out
not to be very sensitive to the details of the radial Regge trajectory. 

Following
Refs.~\cite{PhysRevLett.23.551,Frampton:1969ry,Dominguez:2001zu,Dominguez:2008gg},
we consider the function
\begin{eqnarray}
f_b(t)=\frac{B ( b -1 , \frac{M^2 - t}{a} )}{B ( b -1 , \frac{M^2}{a} )}, \label{f}
\end{eqnarray} 
with $B(x,y)= \Gamma(x) \Gamma(y) /\Gamma(x+y)$ denoting the Euler Beta function.
The function (\ref{f}) fulfills the normalization condition
\begin{eqnarray}
f_b(0)=1.
\end{eqnarray}  
For  $x \gg y $  one has  $B(x,y) \sim \Gamma(y ) x^{-y}$, hence in the 
asymptotic region of $ M^2 - t \gg (b-1) a$ we find
\begin{eqnarray}
f_b(t) \sim \frac{\Gamma\left( \frac{M^2}a +b-1 \right)}{\Gamma\left(
\frac{M^2}a\right)} \left( \frac{a}{M^2-t} \right)^{b-1}.
\end{eqnarray} 
Moreover, this function is positive on the real axis, $t<0$, and has single
poles at $t=M_n^2=M^2 + a n $, with residua read off from the expansion
\begin{eqnarray}
f_b(t) &=&\frac{a}{B ( b -1 , \frac{M^2}{a} )} \label{beta} \\ &\times& \sum_{n=0}^\infty
\frac{\Gamma(2-b+n)}{\Gamma(n+1) \Gamma(2-b)} \frac1{a n + M^2 -t}. \nonumber
\end{eqnarray} 
The function depends on three parameters: the lowest-lying meson mass, $M$,
the string tension, $\sigma = a /(2 \pi)$, and the asymptotic fall-off parameter, $b$.
An interesting feature is the fact that for non-integer values of $b$ a large-$t$
expansion in powers of $1/t$ has zero coefficients.  For integer $b=N+1$
the formula corresponds to exactly $N$ mesons,%
\begin{eqnarray}
f_{N+1}(t) = \prod_{n=0}^N \left(\frac{M^2+a n}{M^2+a n-t}\right). 
\end{eqnarray} 
Particular expressions
corresponding to $b=2,3,4$ are~%
\footnote{One might think that taking $N \to \infty $ the general result would
be recovered, but according to the product formula for the Gamma
function,
\begin{eqnarray}
\Gamma(z)= \lim_{N \to \infty} N^z \prod_{k=1}^N \frac{k}{(k+z)}, \nonumber
\end{eqnarray} 
we see that this is not the case, since 
\begin{eqnarray}
\lim_{N \to \infty} \prod_{n=1}^N \frac{M^2 + a n}{M^2 + an -t} = \lim_{N \to
\infty} N^{t/a} \frac{\Gamma ((M^2 -t)/a)}{\Gamma(M^2/a)} \nonumber
\end{eqnarray} 
and the result is ambiguous. This function has the poles located at
the same place as in (\ref{f}). The ambiguity is manifest in the
choice of the parameter $b$. Generally speaking, the result for
non-integer $N < b< N+1$ has infinitely many resonances but it is
closer to the case of finite $N$ rather than to $N\to\infty$. This
suggests that a method based on truncating the tower of mesons is not
expected to be convergent for increasing $t$ values.}
\begin{eqnarray}
f_2(t) &=& \frac{M^2}{M^2-t},  \\ 
f_3(t) &=& \frac{M^2 (M^2 + a)}{(M^2-t)(M^2+a-t)}, \nonumber \\ 
f_4(t) &=& \frac{M^2 (M^2 + a)(M^2 + 2 a)}{(M^2-t)(M^2+a-t)(M^2+2a-t)}. \nonumber 
\end{eqnarray} 
At asymptotic values of $Q^2$ Eq.~(\ref{beta}) yields
\begin{eqnarray}
f_b(t=-Q^2) = \frac{\Gamma (\frac{M^2}{a} +b-1)}{\Gamma( \frac{M^2}{a})}
\left( \frac{a}{Q^2}\right)^{b-1}. \label{asympt}
\end{eqnarray} 
Thus, the value of $b$ controls the asymptotic fall-off in the $Q^2$ variable. 

\section{From powers to logarithms} 

In this Section we carry out the construction of the pion
electromagnetic form factor which complies to the asymptotic pQCD
constraints. For the pion charge form factor one has the leading power
behavior,
\begin{eqnarray}
F (Q^2) = \frac{16 \pi f_\pi^2 \alpha(Q^2)}{Q^2} \sum_{n=0}^\infty c_n \, \alpha(Q^2)^n, \label{all}
\end{eqnarray} 
with the coefficients $c_n$ calculable in pQCD
(albeit this perturbative series may diverge). We have at LO 
$c_0=1$ 
which is stable in the large-$N_c$ limit, as $\alpha \sim 1/N_c$ and $f_\pi \sim \sqrt{N_c}$. 
For large momenta $F(Q^2)$ is bounded as follows:
\begin{eqnarray}
\frac{C}{Q^4} < F (Q^2) < \frac{C'}{Q^2}.
\end{eqnarray} 
Thus, according to (\ref{asympt}), the admissible possible power
dependence effectively corresponds to $2 < b < 3$.  A fit to the data
yields with $b =2.3(1)$~\cite{Dominguez:2008gg} in agreement with the
above expectations. This is a remarkable result, for it indicates on
the one hand that even at energies where pQCD does not clearly set in,
there seems to be some indirect information on the best possible
fractional power behavior. On the other hand, note that in order to
have a fractional power from the leading twist pQCD result (\ref{all})
we need a non-analytic dependence of the form $e^{-c/\alpha (Q^2)}
\sim (Q^2/ \Lambda^2)^{- 4 \pi c/ \beta_0} $ which is clearly out of
reach for standard perturbation theory. The previous considerations
suggest that pQCD and large $N_c$ Regge models are mutually
incompatible. As we discuss shortly this is not necessarily so.

Now we come to the core of our construction. In
order to generate the asymptotic dependence (\ref{all}) we superpose
the Regge model formula (\ref{f}) over the values of $b$,
\begin{eqnarray}
F(t) = \int_{2}^\infty d b\, \rho(b)  f_b(t), \label{intb}
\end{eqnarray} 
where the density is given by 
\begin{eqnarray}
\rho(b)=\rho_{\rm high}(b)+\rho_{\rm low}(b),
\end{eqnarray}
and
\begin{eqnarray}
\rho_{\rm high}(b) &=& \frac{4\pi}{\beta_0} \frac{16 \pi f_\pi^2}{\Lambda^2}
\left(\frac{a}{\Lambda^2} \right)^{1-b} \frac {\Gamma\left(
\frac{M^2}a\right)}{\Gamma\left( \frac{M^2}a +b-1 \right)} \times \nonumber \\
&& \sum_{n=0}^\infty \frac{c_n}{n!} \left ( \frac{4\pi}{\beta_0}\right )^n (b-2)^n.
\label{high}
\end{eqnarray} 
The coefficients $c_n$ are precisely the same as in Eq.~(\ref{all}).
Note that Eq.~(\ref{high}) corresponds to a Borel transformation of
the original perturbative series, a feature which is welcome in view
that the pQCD series is generally believed to be divergent but
Borel-summable (see, {\em e.g.}, Ref.~\cite{Beneke:1998ui} and
references therein). The formula
\begin{eqnarray}
\int_0^\infty d \epsilon \, \epsilon^n \, x^{-\epsilon} =
\frac{n!}{\log^{n+1} x}
\end{eqnarray} 
is the key ingredient in the {\em power-to-log transmutation}, where
$\epsilon=b-2$.  Note that by taking the spectral density (\ref{high})
we get the right pQCD asymptotics when the large-$Q^2$ behavior of the
Regge model is used.  The lower limit of integration in
Eq.~(\ref{intb}) controls the power of $Q^2$ in front of the RHS of
Eq.~(\ref{all}). In fact, it is the behavior of $\rho(b)$ in the
vicinity of $b=2$ that determines the asymptotic behavior of $F(Q^2)$,
thus $\rho(b)$ is not determined uniquely away from $b=2$.  One could
attempt to use the form (\ref{high}) for all values of $b$.  However,
according to the charge conservation we have to fulfill the
normalization condition
\begin{eqnarray}
F(0) = \int_0^\infty d b \, \rho(b) = 1.
\end{eqnarray} 
Fixing the scale $\Lambda_{\rm QCD} = 250~{\rm MeV}$, we get both at LO and NLO  
\begin{eqnarray}
Z_{\rm high} = \int_2^\infty \, db \, \rho_{\rm high} (b) < 1.
\end{eqnarray} 
To account for the missing strength we add an extra non-perturbative
contribution, $\rho_{\rm low} (b)$, which has support away from
$b=2$. For simplicity is taken in the form of a delta function,
\begin{eqnarray}
\rho_{\rm low} (b) &=& (1-Z_{\rm high}) \delta (b-b_0),
\end{eqnarray} 
with $b_0 = 2.3$, as in the fit of Ref.~\cite{Dominguez:2008gg},
although other less singular distributions could also be
used. Certainly, the presence of $\rho_{\rm low} (b)$ is not affecting
the asymptotics of $F(Q^2)$, which is governed by the behavior near
$b=2$, but it modifies $F(Q^2)$ at lower momenta.

\begin{figure}[tb]
\includegraphics[width=8cm]{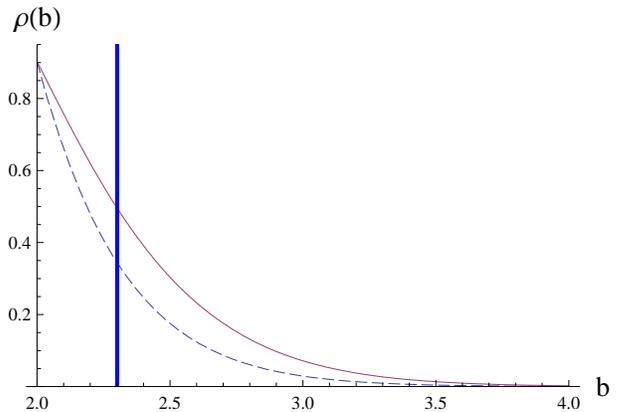} 
\caption{The density $\rho^{\rm high}(b)$ at LO (dashed line) and NLO (solid line). The $\rho^{\rm low}(b)$
contribution is represented by the vertical line at $b=b_0=2.3$.
\label{fig:rho}}
\end{figure}

Explicitly, at LO and NLO we use
\begin{eqnarray}
\rho_{\rm high}^{\rm LO}(b) &=& \frac{4\pi}{\beta_0} \frac{16 \pi f_\pi^2}{\Lambda^2}
\left(\frac{a}{\Lambda^2} \right)^{1-b} \frac {\Gamma\left(
\frac{M^2}a\right)}{\Gamma\left( \frac{M^2}a +b-1 \right)}, \nonumber \\
\rho_{\rm high}^{\rm NLO}(b) &=&\rho_{\rm high}^{\rm LO}(b)\left [ 1+\frac{4\pi}{\beta_0} \frac{6.58}{\pi} \frac{(b-2)}{2!} \right ]
\label{highLONLO}
\end{eqnarray} 
which with parameters (\ref{param}) and $N_f=3$ yield
\begin{eqnarray}
Z_{\rm high}^{\rm LO}=0.27, \;\;\; Z_{\rm high}^{\rm NLO}=0.39.
\end{eqnarray}
The spectral densities (\ref{highLONLO}) are plotted in
Fig.~\ref{fig:rho}, with the dashed and solid lines representing the
LO and NLO formulas, respectively. The $\rho^{\rm low}(b)$
contribution is represented by the vertical line at $b=b_0=2.3$.  We
note that the strength of the spectral density is practically
contained in the interval between 2 and 3, and at large $b$ we have a
very fast fall-off, $\rho_{\rm high} (b) \sim b^{3/2 - M^2/a}
(a/\Lambda^2)^{- b} e^{-b \log(b/e)}$. More generally, we might also
include a finite upper limit of integration using the formula
\begin{eqnarray}
&& \int_{b_1}^{b_2} d b \left( \frac{M_V^2-t}{a} \right)^{1-b} =
\frac{1}{\log (M_V^2-t)/a}  \nonumber \\ 
&& \times\Big[ \left( \frac{M_V^2-t}{a} \right)^{1-b_1}- \left( \frac{M_V^2-t}{a}
\right)^{1-b_2} \Big].
\end{eqnarray} 
The resulting value for $Z_{\rm high}$ changes by a small amount for
$b_1>3$, depending on its precise value.  Actually, by extending the
integration to infinity we are maximizing the impact of perturbative
corrections, and as we see, they are not large.  Thus not much change
is expected from cutting off the integral above $b=3$.

\section{Pole-residue assignment}

\begin{figure}[tb]
\hspace{-12mm}\includegraphics[width=0.53\textwidth]{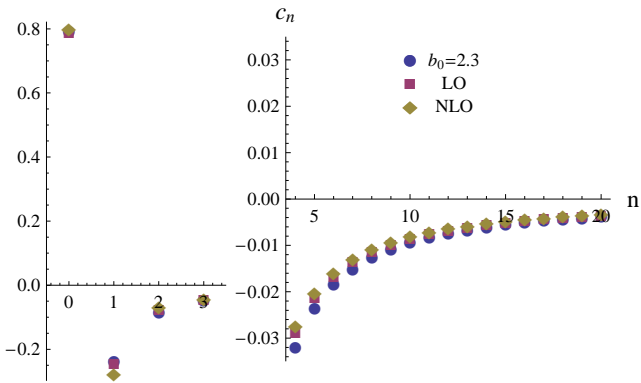} 
\caption{Residues $c_n = F_V(n)g_{V\pi\pi}(n)/M_V(n)$  for the Regge poles of the
pion charge form factor at $M_V(n)^2 = M^2 + a n$ for $n=0,1,2,3$
(left) and for $n =4, 5, \dots $ (right). 
\label{fig:resid}}
\end{figure}

Our procedure is equivalent to imposing pQCD constraints for the
pole-residue assignment in the spectral representation of the pion
charge form factor. Looking formally at the problem, we need to form
the spectral density (\ref{ffv}) in such a manner that the asymptotic
pQCD constraints are satisfied (apart for other constraints, such as
normalization). In the preceding section we have demonstrated
explicitly that it is possible to accomplish this goal.  More
generally, in the large-$N_c$ Regge model we have to choose the
location of poles and fix their residues. Admittedly, there is a
redundancy between shifting the poles or the residues. We decide to
keep the poles fixed at their original location (\ref{reggep}) because
they are phenomenologically well described by the Regge trajectories.
For the residues the prescription of the previous section is equivalent to taking
\begin{eqnarray}
c_n = \int_{2}^\infty db \, \rho(b) \frac{a}{B ( b -1 , \frac{M^2}{a} )} \frac{\Gamma(2-b+n)}{\Gamma(n+1) \Gamma(2-b)}. 
\end{eqnarray}
In Fig.~\ref{fig:resid} we show the values of $c_n$ for the three
considered models: the model with fixed $b$ of Dominguez
\cite{Dominguez:2008gg}, with $\rho(b)=\delta(b-2.3)$ (circles), and
our model for the LO (squares) and NLO (diamonds) cases. We note a
strong similarity between all cases. In particular, the first residue,
$c_0$, is positive and the remaining residues are negative, which
leads to the desired cancellation. At very large values of $n$ (not
displayed) the LO and NLO residues have a larger magnitude than for
the model with fixed $b$. Despite this similarity, we note that our LO
and NLO models do satisfy the asymptotic pQCD constraints, while the
fixed-$b$ model does not. This reflects the subtlety of the
cancellation in the power-to-log transmutation mechanism. We stress
that within our scheme we may achieve the goal of reproducing pQCD
without modifying the spectrum; our spectral method features an
effective way of implementing QCD radiative corrections by appealing
to a modification of the meson wave functions.

At this point it is also interesting to display the values of the
resulting $g_{V\pi\pi}(n)$ couplings. This requires some knowledge on
the vector meson-photon coupling $F_V(n)$. As mentioned above,
quark-hadron duality for large $t$ at the level of the two point
vector correlator requires the Regge model -- parton model matching
condition, Eq.~(\ref{eq:regge-dens}), which for the mass formula,
Eq.~(\ref{reggep}), becomes
\begin{eqnarray}
2 \pi \sigma = 24 \pi^2 F_V^2 / N_c. 
\label{eq:fvsig}
\end{eqnarray} 
This formula works reasonably well already for the lowest $ \rho(770)$
state, where $F_\rho= 150 {\rm MeV}$ yields $\sqrt{\sigma}=530 {\rm
MeV}$, while we expect $\sigma = 420-500 {\rm
MeV}$~\cite{Anisovich:2000kxa}. Following previous
works~\cite{RuizArriola:2006gq,Arriola:2006sv} the formula
(\ref{eq:fvsig}) will be assumed to be valid for all $n$ disregarding
possible non-linearities which are not very relevant within the
present context~\footnote{The sensitivity to details of the Regge
trajectory depends on the computed observable. While for the pion form
factor analyzed here there is some freedom, condensates with proper signs are
crucially dependent on these details, as shown in
Ref.~\cite{RuizArriola:2006gq,Arriola:2006sv}.}.  
With $F_V=150~{\rm MeV}$ and $c_n$ from Fig.~\ref{fig:resid} with
Eq.~(\ref{ffv}) we get
\begin{eqnarray}
g_{\rho\pi\pi} &=& \, \, 4.3 (4.4), \nonumber \\
g_{\rho'\pi\pi} &=& -2.3 (-2.6), \nonumber \\
g_{\rho''\pi\pi} &=& -0.6 (-0.6), \nonumber \\
g_{\rho'''\pi\pi} &=& -0.4 (-0.4), 
\end{eqnarray}
where the first values are for the LO model, and the values in
parenthesis for the NLO model.

\section{Pion charge form factor results}

\begin{figure}[tbc]
\includegraphics[width=8cm]{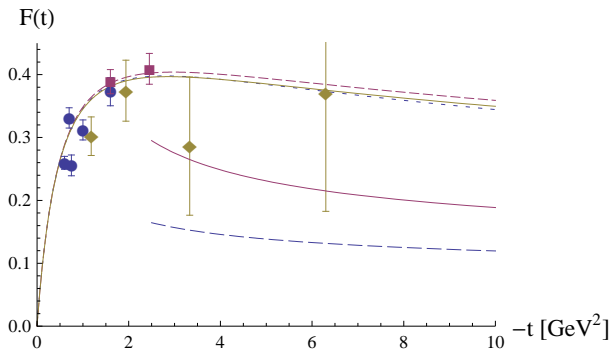}
\caption{The pion charge form factor at
LO and NLO in the space-like region $t <0$: NLO (solid), LO (dashed), and $b_0=2.3$ (dotted). We plot $-t F(t)$ in
the region up to $t=-10 {\rm GeV}^2$ and compare to the TJLAB \cite{Volmer:2000ek,Tadevosyan:2007yd,Horn:2006tm}
(circles and squares) and Cornell \cite{Bebek:1977pe} (diamonds) 
data. The two lower curves 
correspond to the NLO (solid) and (LO) asymptotic pQCD results.
\label{fig:ftpQCD}}
\end{figure}

\begin{figure}[tbc]
\includegraphics[width=8cm]{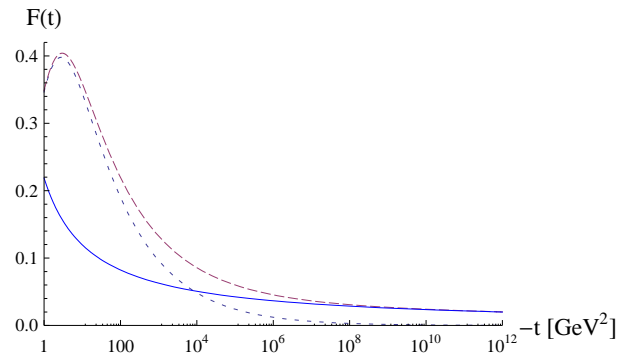} 
\caption{The pion charge form factor in the LO model 
in the space-like region $t <0$. We plot $-t F(t)$ in the region up to very high $t=-10^{12} {\rm GeV}^2$ (on a log
scale). The solid line represents the asymptotic LO pQCD
result. The dashed line is the LO model. The short-dashed line is the model with
$b_0=2.3$. 
\label{fig:fthighQ}}
\end{figure}

\begin{figure}[tbc]
\includegraphics[width=8cm]{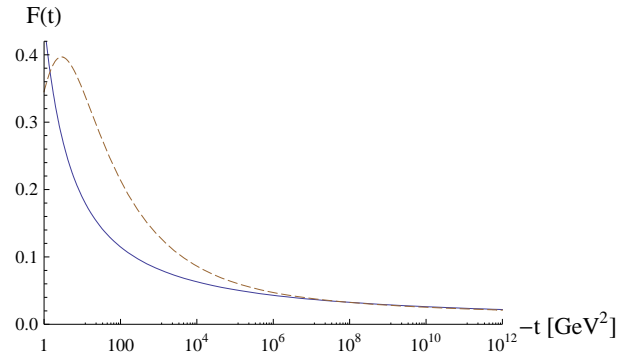} 
\caption{Same as Fig.~\ref{fig:fthighQ} for the NLO case.
\label{fig:fthighQNLO}}
\end{figure}

In Fig.~\ref{fig:ftpQCD} we display the results of our model for the
pion charge form factor and compare them to the TJLAB
\cite{Volmer:2000ek,Tadevosyan:2007yd,Horn:2006tm} and Cornell
\cite{Bebek:1977pe} data. The three lines close to one another and to
the data are the NLO model (solid line), the LO model (dashed line),
and the model with fixed $b=b_0=2.3$. The two lower curves correspond
to the NLO (solid) and (LO) asymptotic pQCD results. We note that in
the range of momenta accessible to experiments all the considered
models yield very close predictions and describe the data well.  These
predictions depart from one another at very high values of $Q^2$, as
can be seen from Fig.~\ref{fig:fthighQ}, where we plot the LO result
(dashed line) and the asymptotic LO pQCD expression (solid line). We
note that the curves meet at $Q^2\sim 10^8~{\rm GeV}^2$, which is a
very high scale. For comparison we also plot the result of the fixed
$b$ model (dotted line), which with the chosen value $b_0=2.3$ decays
as $(1/Q^2)^{1.3}$.  Figure~\ref{fig:fthighQNLO} shows the same study
for the NLO calculation, with the model denoted by the dashed, and the
NLO pQCD calculation by the solid lines, respectively. The two curves
meet at somewhat lower scales, $Q^2\sim 10^7~{\rm GeV}^2$, than for
the LO case of Fig.~\ref{fig:fthighQ}.

\section{Conclusions}

There have been countless attempts to understand the delayed onset of
pQCD in the pion charge form factor. The standard VMD model is known
to fit the available data remarkably well, but shows no obvious link
to pQCD. In the present paper we have approached the problem from the
viewpoint of the large-$N_c$ Regge models.  Our approach exploits
explicitly the quark-hadron duality at a non-perturbative level and
has the genuine advantage that much of the discussion can be carried
out without an explicit reference to the light-cone wave functions
and/or parton distribution amplitudes; many uncertainties in current
calculations seem related to our lack of the detailed knowledge of
these non-perturbative objects used in the description of exclusive
processes. Our generalized VMD model includes infinitely many
resonances, describes the data and simultaneously complies to the
known short-distance pQCD constraints. The present framework requires
a suitable modeling both of the spectrum and the vector meson coupling
to the electromagnetic current. While it describes the so far
experimentally explored space-like momentum region, it is rather hard
to provide estimates of the systematic error of the calculation. The
important feature which has been clearly identified several times in
the past in the analysis of the data is that at large $Q^2$ the pion
form factor seems to have a non-integer power fall-off, which actually
turns out to be in the expected range for the best possible pQCD
power-log behavior, but still is qualitatively different from the
theoretical expectations based on pQCD.  We have shown that there is
no contradiction between both behaviors. Actually, we have spelled out
a simple mechanism where a suitable superposition of non-integer power
fall-offs may transmute into the desired asymptotic pQCD behavior,
including the presence of logarithms.  We have shown that such a
procedure does not spoil the good agreement in the so far
experimentally accessible region down the low energy region where
chiral corrections cause sizable distortions from any large-$N_c$
calculation. Moreover, we are able to reproduce simultaneously the
high-energy pQCD behavior, providing some confidence on the range
where pQCD sets in.  We find that about 1/4 for the LO and about 1/3
for the NLO case of the pion charge is due to the high-energy pQCD
tail in our approach. Finally, the present calculations suggest that
non-perturbative contributions dominate in the region corresponding to
the present and planned experimental data, and would saturate the full
result only at extremely high values, $Q^2\sim 10^7-10^8~{\rm GeV}^2$.


\end{document}